\documentclass{ws-ijmpa}
\usepackage[super,compress]{cite}
\usepackage{graphicx}

\usepackage{amsmath,amssymb,amsfonts}
\usepackage{delarray}
\usepackage[T1]{fontenc}

\newcommand{\Ah}{\ensuremath{\hat{\mathcal A}}}
\newcommand{\An}{\ensuremath{\mathcal A}}
\newcommand{\As}{\ensuremath{{\mathcal A}_\star}}
\newcommand{\C}{\ensuremath{\mathbb C}}

\newcommand{\ds}{\displaystyle}
\newcommand{\g}{\ensuremath{\mathfrak g}}

\newcommand{\Hh}{\ensuremath{\hat{\mathcal{H}}}}
\newcommand{\Hn}{\ensuremath{\mathcal{H}}}
\newcommand{\ka}{\ensuremath{\kappa}}
\newcommand{\la}{\longrightarrow}

\newcommand{\M}{\ensuremath{\mathbb M}}

\newcommand{\Mmn}{\ensuremath{M_{\mu\nu}}}

\newcommand{\mC}{\ensuremath{{\mathcal C}}}
\newcommand{\mD}{\ensuremath{{\mathcal D}}}
\newcommand{\mK}{\ensuremath{{\mathcal K}}}
\newcommand{\mP}{\ensuremath{{\mathcal P}}}

\newcommand{\p}{\ensuremath{\mathfrak p}}

\newcommand{\pmu}{\ensuremath{\partial_{\mu}}}

\newcommand{\R}{\ensuremath{\mathbb R}}

\newcommand{\rhdb}{\ensuremath{\blacktriangleright}}

\newcommand{\so}{\ensuremath{\mathfrak {so}}}

\newcommand{\xmu}{\ensuremath{x_{\mu}}}

\newcommand{\cm}{\'{c}}

\begin{document}

%
\catchline{}{}{}{}{}
%

\title{HERMITIAN REALIZATIONS OF \ka--MINKOWSKI SPACETIME}

\author{DOMAGOJ KOVA\v{C}EVI\'C
}

\address{University of Zagreb, 
Faculty of Electrical Engineering and Computing,\\ 
Unska 3, HR-10000 Zagreb, Croatia\\
domagoj.kovacevic@fer.hr} 

\author{STJEPAN MELJANAC, 
ANDJELO SAMSAROV and ZORAN \v{S}KODA
}
\address{Division for Theoretical Physics, Institute
Rudjer Bo\v{s}kovi\'c, Bijeni\v{c}ka 54, P.O.Box 180, HR-10002
Zagreb, Croatia\\
meljanac@irb.hr, zskoda@irb.hr, asamsarov@irb.hr}

\maketitle

\begin{history}
\received{Day Month Year}
\revised{Day Month Year}
\end{history}

\begin{abstract}
 General realizations, star products and plane waves 
  for \ka--Minkowski spacetime  are considered.
  Systematic construction of general hermitian realization is presented, 
  with special emphasis on noncommutative plane waves 
  and hermitian star product.
  Few examples are elaborated and possible physical applications are mentioned.

\end{abstract}
 

\ccode{PACS numbers: 02.20.Sv, 11.10.Nx, 
AMS Classification: 81R60,16A58, 81T75 }


\section{Introduction}

It has been argued that the description of spacetime as a smooth
manifold is not adequate at the very small distances 
such as those of the order of the Planck length. Instead, in this region, 
the spacetime structure might require a description in terms of
noncommutative geometry. This may result in numerous consequences, some
already seen at the theoretical level,
affecting much of the high energy physics,
including implications for astrophysics. The change in the nature of
the spacetime structure requires the modification of a
mathematical setup and eventually leads to a need of 
rewriting the quantum field theory in the presence 
of quantum group symmetries. 

Since the Poincar\'{e} group describes the symmetry inherent to the
quantum field theory, it is plausible to assume that the symmetry
adequate to such a theory on a noncommutative space
might be described by a quantum deformation 
of the original Poincar\'{e} group. 
One way to accommodate this idea is
within the framework of Hopf algebras.
One of the most extensively studied Hopf algebras in the role of such a
deformation is the \ka-Poincar\'{e} algebra 
\cite{L1,L2,L3,L4,zakrzewski,klm00,DimWess,ACwaves,Govindarajan:2008qa,Govindarajan:2009wt,young,kovacevpoincHopf,poincHopfR,poincHopf}, with $\kappa $
denoting a masslike
deformation parameter, usually associated  with the Planck mass.

It is known for some time that the coproduct of the quantum group 
codifies the curvature in the momentum space and
that a noncommutative spacetime algebra multiplication appears as
dual to the coproduct on the momentum algebra; moreover the 
two dual Hopf algebras together form a noncommutative phase
space algebra built through a cross (smash) 
product algebra construction, more specifically the Heisenberg double  
\cite{Majid:1994cy,KowalskiGlikman:2002we,KowalskiGlikman:2002jr}.
Curiously, the idea that the curvature of momentum space could imply the
noncommutativity of spacetime does not appear to be a recent one,
since it can be traced back to the 
first half of the past century~\cite{snyder}. Later on, this led to a quantum
geometric picture in which curved momentum space is dual to
noncommutative spacetime and where the relation to quantum groups is
made explicit~\cite{majid}.

  The authors of the present paper find $\kappa $-Poincar\'{e} algebra and
  the emergent  $\kappa $-Minkowski spacetime interesting from many 
  reasons. However, the following two may be particularly
  highlighted. Already in the early study of 
  certain limits of quantum gravity coupled to matter 
  in~\cite{AmelinoCamelia:2003xp,Freidel:2003sp,Freidel:2005me}
\footnote{The thesis reported in~\cite{Freidel:2005me} is somewhat
different from the ones 
advocated in~\cite{AmelinoCamelia:2003xp,Freidel:2003sp}
and has been partly challenged by~\cite{schr1,schr2}.}, 
  possible effective quantum field theories on noncommutative 
  $\kappa$-Minkowski space were found
  by integrating out the gravitational degrees of freedom
  from the generating functional. 
  
  The $\kappa$-Minkowski spacetime and $\kappa$-Poincar\'{e} group 
  may also provide a setting appropriate 
  for trapping the signals of quantum gravity effects, 
  what might be an explanation for the observations of
  ultrahigh energy cosmic rays, contradicting the usual understanding
  of electron-positron pair production in collisions of high energy
  photons and other high energy astrophysical processes alike. 
  One possible explanation of the deviations of this
  kind may rely on the modified dispersion relations 
  \cite{ellis,Gambini:1998it,AmelinoCamelia:2000zs,liberati}, whose distortion
  can eventually be traced back to noncommutative structure of the
  spacetime, particularly in the case of $\kappa$ type. All these may be the
  relics of the quantum gravity effects and in turn might allow a description
  in terms of the effective noncommutative quantum field theories (NCQFT). 
  The latter is particularly interesting in light of 
  the arguments supporting the assertion of NCQFT on  
  a $\kappa $-Minkowski space emerging 
  from the quantum gravity in a certain low energy limit 
  \cite{AmelinoCamelia:2003xp}.


     In \cite{mssg} the properties of the star product in the
     {\it{natural}} realization of the $\kappa $-Minkowski algebra 
     are investigated, together with the
     properties of the resulting free scalar field theory constructed
     from this particular star product. The above mentioned realization
     is the type of a differential representation 
     of the $\kappa $-Minkowski algebra
     that is intimately related to the so called
     classical basis of $\kappa $-Poincar\'{e} 
   \cite{KowalskiGlikman:2002we,KowalskiGlikman:2002jr,bp1}.
    These properties have been further studied in \cite{msa11} and the
     analysis has been extended to include the
     implications of the hermitization process on the star product
      and the resulting scalar field theory. 
   In this paper, we extend this analysis to include 
   an arbitrary realization of $\kappa $-Minkowski algebra, 
   investigate the corresponding star products 
   and the effects of hermitization on their properties.
 
  These issues are particularly important 
  when one seeks to find out how the physical results depend 
  on the choice of a realization of the spacetime algebra. 
In this direction, the dependence on the realization has
been studied for the electrodynamics \cite{Harikumar:2011um} 
and the geodesics \cite{Harikumar:2012zi} in $\kappa $-Minkowski spacetime.
Furthermore, the time delay phenomenon for photons having different
  energies has been investigated in the context of relative locality \cite{AmelinoCamelia:2011bm} to see 
how it varies with realizations \cite{Meljanac:2012pv}.
The properties of the scalar field propagation in the
  noncommutative $\phi^4$ model are investigated in
  \cite{Meljanac:2011cs} in the case of the {\it{natural}}
  realization. It would be interesting to compare these results with
  the calculations for other realizations.
While it is desirable to identify the properties  
universal to all realizations of $\kappa $-Minkowski spacetime, those
properties which are not invariant, may, after comparison
with the experimental data, serve to narrow down 
the set of allowed realizations. 
We hope that the experiments may reach 
the necessary level of sensitivity for this in near future (see astrophysical review~\cite{ACpheno}).
To address these issues, it is necessary to set up the
adequate framework and investigate the mathematical properties of the
star products related to each of the realizations of $\kappa $-Minkowski.
The present paper serves this purpose.

 After reviewing in Section 2 the general notions used in the 
   analysis, in Sections 3 and 4 we introduce the star products
   corresponding to various realizations and present their
   properties. Then, in Section 5, we describe the process of hermitization
 and its accompanying impact on the properties of the star product.
 Up to the Section 6 the analysis is general, while in
Section 7 we review the most common concrete examples.
New results are presented in Sections 3, 4 and 5 and the most important
technical tools for understanding these sections are developed in Appendices.    
\section{\ka-Minkowski spacetime and \ka-Poincar\'e algebra}
\label{ksp}

We consider the \ka-Minkowski space $\mathfrak{m}_\ka$, for which the
following commutation relations
\begin{equation}
  [\hat{x}_\mu,\hat{x}_\nu]=i\left(a_\mu\hat{x}_\nu-a_\nu\hat{x}_\mu\right)
  \label{b1}
\end{equation}
are satisfied for $a\in\M^n$.
Latin indices are used for the set $\{1,\ldots,n-1\}$
and Greek indices for the set $\{0,\ldots,n-1\}$. 
The metric of the \ka-Minkowski space is
given by $[\eta_{\mu\nu}]=diag(-1,1,\ldots,1)$.

The Lorentz algebra, $\mathfrak{l}$, is generated by \Mmn\ ($M_{\nu\mu}=
-\Mmn$) satisfying the usual commutation relations
\begin{equation}
  [\Mmn,M_{\lambda\rho}]=M_{\mu\rho}\eta_{\nu\lambda}-M_{\nu\rho}\eta_{\mu
  \lambda}-M_{\mu\lambda}\eta_{\nu\rho}+M_{\nu\lambda}\eta_{\mu\rho}.
  \label{b4}
\end{equation}
The \ka-Minkowski space $\mathfrak{m}_\ka$ can be enlarged to the Lie
algebra $\g_\ka$ which is generated by $M_{\mu\nu}$ and $\hat{x}_\lambda$, 
and where the commutator $[M_{\mu\nu},\hat{x}_\lambda]$ is
\begin{equation}
  [M_{\mu\nu},\hat{x}_\lambda]=\hat{x}_\mu\eta_{\nu\lambda}-\hat{x}_\nu
    \eta_{\mu\lambda}-i\left(a_\mu M_{\nu\lambda}-a_\nu M_{\mu\lambda}\right).
  \label{b7}
\end{equation}
In \cite{mks07}, it is shown that the relation (\ref{b7}) 
is the unique way to equip the direct sum of vector spaces 
$\mathfrak{m}_\ka$ and $\mathfrak{l}$ 
with a Lie algebra structure so that
$\mathfrak{m}_\ka$ and $\mathfrak{l}$ are its Lie subalgebras
which mutually do not commute.

It remains to add the momentum generators $p_0,\ldots,p_{n-1}$ satisfying the
commutation relations
\begin{equation}
  [p_\mu,p_\nu]=0.
  \label{b10}
\end{equation}
Our goal is to enlarge the universal enveloping algebra $U(\mathfrak{g}_\ka)$
by the momentum generators $p_\mu$ to an algebra \Hh\ (see \cite{bp11} for a
similar approach). For the beginning, let us complete the set of commutation
relations for elements \Mmn, $\hat{x}_\mu$ and $p_\mu$:
\begin{equation}
  [p_\mu,\hat{x}_\nu]=-ih_{\mu\nu}(p)
  \label{b13}
\end{equation}
and
\begin{equation}
  [M_{\mu\nu},p_\lambda]=g_{\mu\nu\lambda}(p).
  \label{b16}
\end{equation}
Functions $h_{\mu\nu}$ and $g_{\mu\nu\lambda}$ are real, satisfying
$\lim_{a\rightarrow0}h_{\mu\nu}=\eta_{\mu\nu}\cdot1$,
$\det h\neq0$
and
$\lim_{a\rightarrow0}g_{\mu\nu\lambda}=
p_\mu\eta_{\nu\lambda}-p_\nu\eta_{\mu\lambda}$.
Now, we require that all Jacobi identities are satisfied.
This gives conditions on $h_{\mu\nu}$ and $g_{\mu\nu\lambda}$.
If exactly one of the three generators appearing in
a Jacobi identity is $p_\mu$,
the remaining two generators can either be \Mmn\ and $M_{\lambda\rho}$,
or $\hat{x}_\mu$ and $\hat{x}_\nu$, or \Mmn\ and
$\hat{x}_\lambda$. These three cases lead to the respective 
 system of differential equations:
\begin{eqnarray}
  &&\frac{\partial g_{\lambda\rho\sigma}}{\partial p_\alpha}g_{\mu\nu\alpha}-
      \frac{\partial g_{\mu\nu\sigma}}{\partial p_\alpha}g_{\lambda\rho\alpha}=
      g_{\mu\rho\sigma}\eta_{\nu\lambda}-g_{\nu\rho\sigma}\eta_{\mu\lambda}-
      g_{\mu\lambda\sigma}\eta_{\nu\rho}+g_{\nu\lambda\sigma}\eta_{\mu\rho},
      \label{b19}\\
  &&\frac{\partial h_{\lambda\nu}}{\partial p_\alpha}h_{\alpha\mu}-
      \frac{\partial h_{\lambda\mu}}{\partial p_\alpha}h_{\alpha\nu}=
      a_\mu h_{\lambda\nu}-a_\nu h_{\lambda\mu},
      \label{b22}\\
  &&\frac{\partial g_{\mu\nu\lambda}}{\partial p_\alpha}h_{\alpha\rho}-
      \frac{\partial h_{\lambda\rho}}{\partial p_\alpha}g_{\mu\nu\alpha}=
      h_{\lambda\nu}\eta_{\mu\rho}-h_{\lambda\mu}\eta_{\nu\rho}+
      a_\mu g_{\nu\rho\lambda}-a_\nu g_{\mu\rho\lambda},
      \label{b23}
\end{eqnarray}
putting the constraints on the possible choices of the functions $h_{\mu\nu}$ and $g_{\mu\nu\lambda}$.
For more details see also \cite{mk11}.

The system of differential equations (\ref{b19})-(\ref{b23})
has infinitely many solutions. In Section \ref{ghr}, 
we construct a large family of these solutions
from a particular one.
It is important to mention that functions $g_{\mu\nu\lambda}(p)$ are completely
determined by the functions $h_{\mu\nu}(p)$.
Generators \Mmn\ and $p_\mu$ form \ka-Poincar\'{e} algebra. 
This algebra is denoted by \p.
Here we consider the algebra structure only. 
The link with \ka-Poincar\'{e} algebra
is presented in \cite{km11}.

Let us define the invertible element $Z$ (see \cite{mks07}) 
in \Hh\ by the shift property
\begin{equation}
  [Z,\hat{x}_\mu]=ia_\mu Z,
  \label{b25}
\end{equation}
\begin{equation}
  [Z,p_\mu]=0.
  \label{b28}
\end{equation}
and $\lim_{a\rightarrow0}Z=1$.
Element $Z$ is in \Hh\ and we emphasize it because of his properties.
One can show that (\ref{b25}) and (\ref{b28}) together with the boundary
condition $\lim_{a\rightarrow0}Z=1$ define $Z$ uniquely.
Equations (\ref{b1}) and (\ref{b25}) show that $\hat{x}_\mu$ and $Z$
generate the Lie algebra.

In the next section we present how \Mmn\ can be expressed as a function of
$\hat{x}_\mu$ and $p_\mu$, while $Z$ can be expressed in terms of $p_\mu$.
Hence, \Hh\ is generated  by $\hat{x}_\mu$ and $p_\mu$ only.
To be more precise, \Hh\ is the quotient of the free algebra generated by
$\hat{x}_\mu$ and $p_\mu$  
 {\footnote{\Hh\ should be generated by $\hat{x}_\mu$ and
   $\hat{p}_\mu$, but we simplify the notation by writing
   $\hat{p}_\mu \equiv  p_\mu$. }}
 and the ideal generated by the relations
(\ref{b1}), (\ref{b10}) and (\ref{b13}).
Let us denote by \Ah\  the subalgebra of \Hh\ generated by $\hat{x}_\mu$.
Then  we can introduce the action \rhdb\ of \Hh\ on \Ah\  defined by
\begin{equation}
   \hat{g}(\hat{x}) \rhdb\ 1 =\hat{g}(\hat{x}),\;\; 
\hat{x}_\mu\rhdb\hat{g}(\hat{x})=\hat{x}_\mu\hat{g}(\hat{x}),\;\;
    \hat{g}(\hat{x})\in\Ah,
\end{equation}
\begin{equation}
  p_\mu\rhdb1=0,\;\Mmn\rhdb1=0,
\end{equation}
\begin{equation} 
p_\mu\rhdb\hat{g}(\hat{x})=[p_\mu,\hat{g}(\hat{x})]\rhdb1=
    p_\mu\hat{g}(\hat{x})\rhdb1,\,\,
    \Mmn\rhdb\hat{g}(\hat{x})=[\Mmn,\hat{g}(\hat{x})]\rhdb1=
    \Mmn\hat{g}(\hat{x})\rhdb1.
\end{equation}
For more details on the action \rhdb, see \cite{km11}.

Let us further introduce the anti-involution operator $\dagger$ 
by $\lambda^\dagger=\overline{\lambda}$, 
for $\lambda\in\C$ and bar denoting the ordinary
complex conjugation, $\hat{x}_\mu^\dagger=\hat{x}_\mu$,
$\Mmn^\dagger=-\Mmn$ and $p_\mu^\dagger=p_\mu$. Then $Z^\dagger=Z$.
Since functions $h_{\mu\nu}$ and $g_{\mu\nu\lambda}$ are real and
$[a,b]^\dagger=-[a^\dagger,b^\dagger]\;\forall a,b\in\Hh$,
relations (\ref{b1})-(\ref{b16}) which define the algebra \Hh\ remain unchanged
under the action of $\dagger$.
Also, relations (\ref{b25})-(\ref{b28}) which define $Z$ remain unchanged.
Thus, the commutation relations defining  \Hh\ are invariant under
operation $\dagger$.

\section{General realizations and the star product}
\label{hrsp}

\subsection{General considerations}

Let us consider realizations of \Hh.
We want to express noncommutative coordinates $\hat{x}_\mu$ in
terms of commutative coordinates $x_\mu$ and momenta $p_\mu$.
Hence, let us consider the Weyl algebra \Hn, generated by $x_\mu$ and
$p_\mu$ satisfying
\begin{equation}
  [x_\mu,x_\nu]=[p_\mu,p_\nu]=0
\end{equation}
and
\begin{equation}
  [p_\mu,x_\nu]=-i\eta_{\mu\nu} \cdot 1.
\end{equation}
Generators $x_\mu$ and $p_\mu$ satisfy
$x_\mu^\dagger=x_\mu$ and $p_\mu^\dagger=p_\mu$.
Now, we realize $\hat{x}_\mu(x,p)$ in the form
\begin{equation}\label{c1}
  \hat{x}_\mu=x_\alpha h_{\alpha\mu}(p)+\chi_\mu(p),
\end{equation}
where the constant term in $h_{\alpha\mu}$ is $\delta_{\mu\alpha}$
and $\chi_\mu(0)=0$.
Note that this is somewhat more general form of the differential
representation than those used in~\cite{mks07,km11,mst}, 
where it was assumed that $\chi_\mu=0$. We include the additional term
$\chi_\mu$ in order to construct hermitian realizations.

The relation~(\ref{b1}), upon substitution~(\ref{c1}), 
gives a differential equation for $\chi$:
\begin{equation}
  \frac{\partial\chi_\nu}{\partial p_\alpha} h_{\alpha\mu}-
  \frac{\partial\chi_\mu}{\partial p_\alpha} h_{\alpha\nu}=
  a_\mu\chi_\nu-a_\nu\chi_\mu.
\end{equation}
For a given $h_{\mu\nu}$, this differential equation has infinitely many
solutions, each of which 
generates a realization, which is in general nonhermitian.
If we require that $\hat{x}_\mu$ be the hermitian operator, then again
there will be an infinite subfamily of solutions. In Section~\ref{ghr}, 
it is shown how to obtain a large class of
(hermitian) solutions from one particular solution.

We further introduce the angular momentum operator
 \Mmn\ in the following way. Denote by $P_{\mu}$ a momentum operator
 $p_{\mu}$, Eq.(\ref{b13}), satisfying the particular commutation relation, namely  
  $[P_\mu,\hat{x}_\nu] = -i (Z^{-1}
 {\eta}_{\mu\nu} - a_{\mu}P_{\nu}) $. By comparison with the Eq.(\ref{b13}),
this choice for the commutator corresponds to choosing the function 
$h_{\mu\nu} = Z^{-1}{\eta}_{\mu\nu} - a_{\mu}P_{\nu} $.
Then the angular momentum operator can be introduced as
\begin{equation}
  \Mmn=i\left(\hat{x}_\mu P_\nu-\hat{x}_\nu P_\mu\right)Z
\end{equation}
(see also~\cite{mssg,bp1,mks07}). It is easy to check that
$\Mmn^\dagger=-\Mmn$ if $\hat{x}_\mu^\dagger=\hat{x}_\mu$.

Let \An\ be the subalgebra of \Hn\  generated by $x_\mu$. Similarly to \Hh\ and
the action \rhdb, we define the action $\rhd$ of \Hn\ on \An\ by 
\begin{enumerate}
  \item $ g(x) \rhd 1 = g(x),\;\;  \xmu\rhd g(x)=\xmu g(x),\;\;g(x)\in\An$
  \item $p_\mu\rhd g(x)=[p_\mu,g(x)]\rhd1=p_\mu g(x)\rhd1.$
\end{enumerate}
Again, for more details on the action $\rhd$, see \cite{km11}.
Since $g(x)\rhdb1=\hat{g}(\hat{x})$ and $\hat{g}(\hat{x})\rhd1=g(x)$,
for $g(x)\in\An$, the action $\rhd$ is in a certain way the inverse of the
action \rhdb. This relationship between the two operations is a natural
consequence of the fact that the
algebras \Ah\ and \An\ are isomorphic as the vector spaces. Moreover,
they are also isomorphic
 at the algebra level if the pointwise multiplication on \An\ is replaced by the
star product. Let us denote by \As\ the algebra \An\ in which the star product
is used instead of pointwise multiplication.
Then there is a linear map $T:\Ah\la\As$ given by
\begin{equation} \label{isomorphism}
  T(\hat{g}(\hat{x}))=\hat{g}(\hat{x})\rhd 1=:g(x),
\end{equation}
where $g(x)$ is the result.
It can be shown that this map is an isomorphism. 
The essential observation is 
that the difference $\hat{x}_\mu-x_\mu$ is linear or 
higher order in $p_\nu$ and we see by induction on
the order of monomial that $T(\hat{x}_{i_1}\cdots\hat{x}_{i_k})$
equals $x_{i_1}\cdots x_{i_k}$ up to the lower order terms. 
Thus the bijectivity of $T$ easily follows from PBW theorem.
For a detailed discussion of the properties 
of the operator $T$ (in the case $\chi = 0$) 
we refer the reader to Ref.~\cite{mssg}.
In this setting, the star product is defined by
\begin{equation} \label{starproductdefinition}
  (f\star g)(x)=\hat{f}(\hat{x})\hat{g}(\hat{x})\rhd1=\hat{f}(\hat{x})\rhd g(x).
\end{equation}


Suppose we have two arbitrary sets $\{ X_\mu, P_\nu \}$ and $ \{ x_\mu,
p_\nu \}$ of canonical coordinates and momenta, 
each  forming a set of canonical variables,
$[P_\mu,X_\nu] =-i\eta_{\mu\nu},$ and $[p_\mu, x_\nu] =-i\eta_{\mu \nu},$
respectively. Then, we look for a similarity transformation $S$ 
that maps one set to another in the form: 
\begin{eqnarray} \label{umetak}
X_\mu & =& S x_{\mu} S^{-1} = x_{\alpha} \tilde{h}_{\alpha \mu}(p) 
+ \tilde\chi_\mu(p),
\nonumber \\
P_\mu &=& S p_\mu S^{-1} = \Lambda_\mu (p). 
\end{eqnarray}
Note that $\Lambda$ is taken to depend on $p$ only because we want to keep
linearity of $\hat{x}_\mu$ in $x$.
For simplicity, we now assume $\tilde\chi_\mu(p)=0$. Then we can write
\begin{equation}
\hat{x}_\mu  = X_\alpha H_{\alpha \mu} (P)
   = x_\nu \tilde{h}_{\nu \alpha} (p) H_{\alpha \mu} (\Lambda (p))
   = x_\nu {h}_{\nu \mu} (p),
\end{equation}
implying that
\begin{equation}
  {h}_{\nu \mu}(p)=\tilde{h}_{\nu \alpha}(p)H_{\alpha \mu}(\Lambda(p)).
\end{equation}
Since $[P_\mu, X_\nu] =-i\eta_{\mu \nu}, $ after introducing $P_\mu$
and $X_\nu$ from (\ref{umetak}) into this relation, we obtain
\begin{equation} \label{umetak1}
 \frac{\partial\Lambda_\mu}{\partial p_\alpha}
\tilde{h}_{\alpha\nu}(p)={\eta}_{\mu\nu},
\end{equation}
where we have taken into account that the other set of commutative
variables is canonical as well, $[p_\mu,x_\nu] =-i\eta_{\mu \nu}. $
Relation (\ref{umetak1}) then means that the matrix 
$\left(\tilde{h}_{\alpha\nu}\right)$ is the inverse
of the matrix $ \left(\frac{\partial \Lambda_\mu}{\partial p_\alpha}\right)$,
\begin{equation} \label{umetak2}
 \tilde{h}(p)=
{\left(\frac{\partial\Lambda}{\partial p}\right)}^{-1}.
\end{equation}
Thus, the similarity transformations~(\ref{umetak}) allow effective passage
between the realizations. 

\subsection{Plane waves}

It is clear that the differential realization (\ref{c1}) need not
be hermitian in general, in the sense that the condition 
$(\hat{x}_\mu)^\dagger= \hat{x}_\mu$ may fail. 
However, for a given $h_{\mu\nu}(p)$, it may be made
hermitian by cleverly choosing the function $\chi_\mu (p)$. Since for 
each realization of the form~(\ref{c1}) there is a corresponding star product, 
this is the case for the hermitian realizations as well. 
The star products corresponding to hermitian realizations 
are called here the hermitian star products.  
Note that we do not use this term in a different sense that 
the property $\overline{f \star g} = \bar{f} \star \bar{ g}$ holds. 
Indeed, the hermitian star products in our sense do not need to 
satisfy this property in general.
However, for
some very common orderings (realizations) including the
Weyl symmetric and left-right symmetric ordering,
this property is satisfied.
It is also noteworthy that for each $h_{\mu\nu}(p)$ there exists a
whole family of hermitian realizations (infinitely many) 
and consequently there exists a
whole family of star products corresponding to these hermitian
realizations. Each star product is thus characterized by specifying
two functions, $ h_{\mu\nu}(p) $ and  $\chi_\mu (p)$. In what follows 
we therefore use the symbol
   ${\star}_{h,\chi }$ to denote the star product and  also add the
   subscripts $ h $ and  $\chi$
   to various quantities that are to be introduced in the following,
   to reflect their dependence on the particular realization 
in the class (\ref{c1}).
The analysis is first carried out for the most general situation of
an arbitrary $ h_{\mu\nu}(p) $ and  $\chi_\mu (p)$, not necessarily
corresponding to a  hermitian realization,
 and then only after making this case clear, we specialize it to
 the hermitian realization and the hermitian star product.   

Let us define functions ${\mP}_h$ and ${\mC}_{h, \chi}$ by 
(cf. Appendix A for justification)
\begin{equation}\label{c19}
  e^{ik\hat{x}}\rhd e^{iqx}=\mC_{h, \chi}(k,q)e^{i\mP_{h}(k,q)x},
\end{equation}
where $\hat{x}$ on the left hand side is given by (\ref{c1}), with
arbitrary $ h_{\mu\nu}(p) $ and $ \chi_\mu (p)$. 
Similarly, the function ${\mK}_h $ can be defined as the special case of
the above,
\begin{equation}\label{definitionofk}
  e^{ik\hat{x}}\rhd 1=\mC_{h, \chi}(k)e^{i\mK_{h}(k)x},
\end{equation}
taking into account that $\mK_{h}(k) =  \mP_{h}(k,0)$ and $\mC_{h, \chi}(k) = \mC_{h, \chi}(k,0)$.
It is shown in Appendix A that
\begin{equation}\label{codkaku}
 \mC_{h, \chi}(k,q)=e^{i\int_0^1(k\chi(\mP_{h}(tk,q)))\,dt}.
\end{equation}
This expression is valid for any $ h_{\mu\nu}(p) $ and $ \chi_\mu (p)$. While
the dependence of \mC\ on $\chi$ is obvious from (\ref{codkaku}), the
dependence on $h$ enters through the function  ${\mP}_h$.
If $\chi=0$, then $\mC_{h, \chi} (k,q)=1$ (see \cite{mks07,km11,mmss,mmssM,meljSkodaSvrtan} for details).
It is convenient to use this special case when $\chi=0$ to  define the
function ${\mD}_h $ by
\begin{equation}\label{c22}
  e^{ikx} \; {\star}_{h,\chi=0 } \; e^{iqx}=e^{i\mD_{h,\chi=0 }(k,q)x},
\end{equation}
where it is understood that $h$ and $\chi$  refer to the functions $h
=  h_{\mu\nu}(p) $ and $\chi_\mu = \chi_\mu (p)$, respectively,
appearing  in
the Eqs.(\ref{b13}) and (\ref{c1}).
We emphasize here that for a given  $ h_{\mu\nu}(p) $, the function  $\mD_{h,\chi}(k,q)$
is the same, no matter which form $\chi_\mu$ takes. That is,
$\mD_{h,\chi }(k,q) = \mD_{h,\chi=0 }(k,q) \equiv \mD_{h }(k,q) $.
It is related to $  {\mP}_h (k,q)$   by $\mD_{h}(k,q) = \mP_{h}(\mK_{h}^{-1}(k),q)$.
More details can be found in \cite{mssg,msa11,meljSkodaSvrtan}. 

   We can now introduce the noncommutative plane waves
   $\hat{e}_k^{+}(\hat{x})$  with label $+$ by
\begin{equation}\label{ekaplus}
   \hat{e}_k^{+}(\hat{x}) =  
 \frac{1}{ \mC_{h,\chi }({\mK}_h^{-1}(k))} e^{i{\mK}_h^{-1}(k)\hat{x}}.
\end{equation}
Relation (\ref{definitionofk}) then shows that this noncommutative
plane wave can be related to a commutative plane wave $e^{ikx}$
through the isomorphism (\ref{isomorphism}), namely,
\begin{equation}\label{ekaplusaction}
   \hat{e}_k^{+}(\hat{x}) \rhd 1 = e^{ikx},
\end{equation}
showing that $\hat{e}_k^{+}(\hat{x})$ 
corresponds to the commutative plane wave $e^{ikx}$.

The star product corresponding to a generic $ h_{\mu\nu}(p) $ and
$\chi_\mu (p)$ is readily deduced from (\ref{starproductdefinition})  
and (\ref{ekaplusaction}) as
\begin{equation}\label{c31}
  e^{ikx} ~ {\star}_{h,\chi } ~  e^{iqx}=   \frac{1}{ \mC_{h,\chi }({\mK}_h^{-1}(k))}  e^{i\mK_h^{-1}(k)\hat{x}}
      \frac{1}{ \mC_{h,\chi }({\mK}_h^{-1}(q))}  e^{i\mK_h^{-1}(q)\hat{x}}\rhd1.
\end{equation}
Particularly interesting may be the star product between two plane waves
with 'opposite' momenta, i.e. $e^{iS_h (k)x} ~ {\star}_{h,\chi } ~  e^{iqx}$.
Here $k\mapsto S_h (k)$ is the antipode defined in the current context by
$\mD_h (S_h (k), k) = \mD_h (k, S_h (k)) =0. $ We come back to
this particular form of the star product below when considering 
the hermitian realizations and the star products corresponding to them.

For the time being, 
it remains to finalize the calculation for 
$e^{ikx} ~ {\star}_{h,\chi }  ~ e^{iqx}$,
\begin{eqnarray}
  e^{ikx} ~ {\star}_{h,\chi } ~  e^{iqx} &=&  \frac{1}{ \mC_{h,\chi }({\mK}_h^{-1}(k))} \frac{1}{ \mC_{h,\chi }({\mK}_h^{-1}(q))}
  e^{i\mK_h^{-1}(k)\hat{x}} e^{i\mK_h^{-1}(q)\hat{x}} \rhd 1 =  \nonumber\\
  &=&    \frac{1}{ \mC_{h,\chi }({\mK}_h^{-1}(k))} \frac{1}{ \mC_{h,\chi }({\mK}_h^{-1}(q))}    e^{i\mD_s(\mK_h^{-1}(k),\mK_h^{-1}(q))\hat{x}}
        \rhd 1 \nonumber\\
  &=& \frac{\mC_{h,\chi }  (\mD_s(\mK_h^{-1}(k),\mK_h^{-1}(q)))   }{\mC_{h,\chi }({\mK}_h^{-1}(k)) \mC_{h,\chi }({\mK}_h^{-1}(q)) }  
  e^{i \mK_{h}(\mD_s(\mK_h^{-1}(k),\mK_h^{-1}(q)) )     x}.
                      \label{generalstarproducthahi1}\\
  &\stackrel{(\ref{importantidentity})}{=}& \frac{\mC_{h,\chi } (\mD_s(\mK_h^{-1}(k),\mK_h^{-1}(q)))   }{\mC_{h,\chi }({\mK}_h^{-1}(k)) \mC_{h,\chi }({\mK}_h^{-1}(q)) }  
  e^{i \mD_h (k, q) x} \nonumber\\
  &=& \frac{\mC_{h,\chi } (\mK_h^{-1}(\mD_h (k, q) ))   }{\mC_{h,\chi }({\mK}_h^{-1}(k)) \mC_{h,\chi }({\mK}_h^{-1}(q)) }  
  e^{i \mD_h (k, q) x}, \label{generalstarproducthahi}
\end{eqnarray}
where the identity (\ref{importantidentity}) is again applied in the
numerator to simplify the argument of the \mC ~ factor.

If we restrict ourselves to the class of realizations where the plane
waves $e^{ikx}$ and  $e^{iS_h(k)x}$ are orthogonal and normalized, that is
$$
 \ds \int d^nx e^{iS_h(k)x} ~ {\star} ~ e^{iqx}=
{(2\pi)}^n  {\delta}^{(n)}(k-q),
$$
than the factors \mC ~ are  calculated in a particularly simple way as
(see Appendix C  for details)
\begin{equation} \label{determinant}
  \mC_{h,\chi}({\mK}_h^{-1}(k))
   =\frac{1}{\sqrt{\left|det\left(\frac{\partial(\mD_h
  (S_h (k),q))_\mu}
  {\partial q_\nu}\right)_{k=q}\right|}}.
\end{equation}
Otherwise, it is calculated as a special case 
of the more general factor (\ref{codkaku})
by setting $q = 0, ~$ i.e.  $\mC_{h, \chi}(k) = \mC_{h, \chi}(k,0)$.  
   In~\cite{msa11}, the factor $\mC_{h, \chi} ({\mK}_h^{-1}(k))$
has been calculated with the help of Eq.(\ref{determinant}),
for one special hermitian version of the {\it{natural}} realization 
and it can be shown
that this result is consistent with that obtained through Eq.(\ref{codkaku}).
We recapitulate this result among other results related to the
{\it{natural}} realization in Section 5.

\section{Hermitian realizations}

  We now turn to the problem of hermitization of the realizations
  considered so far.
One way to obtain the hermitian realization is to start from the nonhermitian
realization for which $\chi=0$ and then consider
$\hat{x}_\mu^{(\frac12)}=\frac12(\hat{x}_\mu+\hat{x}_\mu^\dagger)$ (see
\cite{msa11,mst}). The operators $\hat{x}^{(\frac12)}$ 
are hermitian and it remains to prove that they satisfy (\ref{b1}). 
Actually, the operators $\hat{x}_\mu^{(\beta)}$ of the
form $\hat{x}_\mu^{(\beta)}=\beta\hat{x}_\mu+(1-\beta)\hat{x}_\mu^\dagger$
satisfy the relation (\ref{b1}). The left hand side of (\ref{b1}) has the form
\begin{eqnarray}\label{c13}
  [\hat{x}_\mu^{(\beta)},\hat{x}_\nu^{(\beta)}]=\beta^2[\hat{x}_\mu,\hat{x}_\nu]
    +\beta(1-\beta)\left([\hat{x}_\mu,\hat{x}_\nu^\dagger]+[\hat{x}_\mu^\dagger,
    \hat{x}_\nu]\right)+(1-\beta)^2[\hat{x}_\mu^\dagger,\hat{x}_\nu^\dagger].
\end{eqnarray}
The important observation is the identity $[\hat{x}_\mu,\hat{x}_\nu^\dagger]+
[\hat{x}_\mu^\dagger,\hat{x}_\nu]=[\hat{x}_\mu,\hat{x}_\nu]+
[\hat{x}_\mu^\dagger,\hat{x}_\nu^\dagger]$.
Hence, (\ref{c13}) transforms to
\begin{eqnarray}
  [\hat{x}_\mu^{(\beta)},\hat{x}_\nu^{(\beta)}]=\beta[\hat{x}_\mu,\hat{x}_\nu]
    +(1-\beta)[\hat{x}_\mu^\dagger,\hat{x}_\nu^\dagger].
\end{eqnarray}
Since $\hat{x}_\mu$ and $\hat{x}_\mu^\dagger$ satisfy (\ref{b1}),
$\hat{x}_\mu^{(\beta)}$ also satisfies (\ref{b1}).

For the nonhermitian realization $\hat{x}_\mu=x_\alpha h_{\alpha\mu}(p)$,
the corresponding hermitian realization $\hat{x}_\mu^{(\frac12)}$ has the form
\begin{equation}\label{c15}
  \hat{x}_\mu^{(\frac12)}=x_\alpha h_{\alpha\mu}(p)+
  \frac12[h_{\alpha\mu}(p),x_\alpha].
\end{equation}
Since $[h_{\alpha\mu},x_\alpha]=-i\frac{\partial h_{\alpha\mu}}
{\partial p_\alpha}$, (\ref{c15}) transforms to
\begin{equation}\label{c16}
  \hat{x}_\mu^{(\frac12)}=\hat{x}_\mu-\frac i2\frac{\partial h_{\alpha\mu}}
   {\partial p_\alpha}.
\end{equation}
The last expression is more convenient for calculation.

Now, let us consider few examples that can be hermitized.
A large class of {\itshape covariant}
\footnote{We use the term {\itshape covariant} in a sense that we treat time
  and space components on equal footing so that we are able to write all
  coordinates in a form of a Lorentz/$\mathrm{GL}(n)$ vector. 
  Consequently, all expressions are
  written in terms of Lorentz/$\mathrm{GL}(n)$ tensors.}
realizations can be found in \cite{mks07,km11}. The
{\itshape noncovariant} realizations are analyzed in \cite{mk11}. Let $A=-(ap)$
and $B=-a^2p^2$.
For the {\itshape natural} realization $\ds \hat{x}_\mu=X_\mu Z^{-1}-(aX)P_\mu$
and $\ds \hat{x}_\mu^{(\frac12)}=\hat{x}_\mu+\frac i2\frac{a^2P_\mu}{\sqrt{1
-B}}$ where $Z^{-1}=-A+\sqrt{1-B}$.
For the {\itshape left covariant} realization $\ds \hat{x}_\mu=x_\mu(1-A)$
and $\ds \hat{x}_\mu^{(\frac12)}=\hat{x}_\mu+\frac i2a_\mu$.
For the {\itshape right covariant} realization $\ds \hat{x}_\mu=x_\mu-a_\mu(xp)$
and $\ds \hat{x}_\mu^{(\frac12)}=\hat{x}_\mu+\frac{ni}2a_\mu$, with
$n$ being the spacetime dimension.
Finally, for the  realization $\ds \hat{x}_\mu=x_\mu-(ax)p_\mu -a_\mu(xp),$
 we have $\ds \hat{x}_\mu^{(\frac12)}=\hat{x}_\mu+\frac{n+1}2ia_\mu$.

Let us mention that in the class of realizations with $\chi = 0$ there exists
one which is hermitian (see also~\cite{bp2}). 
This one corresponds to the case when
$[x, h_{\mu \nu}(p)]= 0$ and  is determined by setting
\begin{equation}
  \frac{\partial h_{\alpha\mu}}{\partial p_\alpha}=0.
\end{equation}

\section{Noncommutative plane waves in the hermitian realization and 
the hermitian  star product}

 In Section 3 the noncommutative plane wave 
 $\hat{e}^+_k (\hat{x})$ has been introduced,
 which, through the isomorphism~(\ref{isomorphism}), 
 corresponds to the ordinary commutative plane wave $e^{ikx}$.
 Analogously, one can ask for the form of the noncommutative plane wave 
 that corresponds to the ordinary commutative plane
 wave through the same isomorphism
 (\ref{isomorphism}), but this time
 with the 'opposite' or 'inverse' momentum, $e^{iS(k)x}$. 
The noncommutative plane wave with this property we label 
with superscript $-$, thus writing $\hat{e}^-_k (\hat{x})$.
To answer the foregoing question and to find the required correspondence, 
it is necessary to work with the hermitian realizations. 
In the case that the realization for $\hat{x}$ is hermitian, 
the noncommutative plane wave $\hat{e}^+_k (\hat{x})$ 
is a unitary operator, so that the hermitian conjugation operation 
(anti-involution) $\dagger $ is well defined when applied to it.
  Proceeding along these lines and using 
the identity~(\ref{josjedanvazanidentitet1}), 
it is straightforward to establish
 that the noncommutative plane wave $\hat{e}^-_k (\hat{x})$,
 may be introduced via the formula 
\begin{equation}\label{ekaminus}
   \hat{e}_k^{-}(\hat{x}) =  
 \frac{\mC_{h,\chi }({\mK}_h^{-1}(k))}{ \mC_{h,\chi
 }({\mK}_h^{-1}(S_h (k)))}  {({\hat{e}_k^{+}(\hat{x})})}^{\dagger}, \quad
 \hat{e}_k^{-}(\hat{x}) \rhd 1  = e^{i S_h (k) x}.
\end{equation}
With this and~(\ref{ekaplus}), we have defined a complete set of
noncommutative plane waves, appearing in the decomposition of any
element in \Ah\ . They correspond to the plane waves on the commutative space,
$e^{ikx}$, i.e. $e^{iS(k)x}$, to which any element of \As\ is reduced
when shattered into pieces.

Before we proceed and decompose generic elements 
of \As\ and \Ah\  into their respective Fourier modes, 
we need to introduce a generalization of the ordinary complex conjugation,
which serves to define a scalar product on \As\ in a similar way as
the ordinary complex conjugation (designated by the bar) serves to
define the usual scalar product on \An, namely
$(f,g)=\int\mathrm{d}^n x\overline{f}g$. It should be noted that the
operation $\dagger $ introduced on $\Hh$ in Section 2 
is the adjoint operator 
with respect to this scalar product, $(f,A g) = (A^\dagger f, g)$.
Namely, the generators $\hat{x}_\mu$ and $p_\mu$ of $\Hh$  
can be viewed as operators via their action on \An\ and their adjoints
$\hat{x}_\mu^\dagger$ and $p_\mu^\dagger$ considered in the previous sections 
were calculated with respect to this scalar product on \An. 
Beside that the generalized complex conjugation
operation $* $ is used to define a scalar product on \As\
\begin{equation} \label{starproductonastar}
  (f,g)_\kappa = \int\mathrm{d}^nxf^*\star g,
\end{equation}
for any $f,g\in\As$. It is also required to have 
a smooth limit to the ordinary bar, when
the deformation parameter goes to zero. The operation which is the adjoint operation with respect
to the scalar product (\ref{starproductonastar})  we label by
$\ddagger$. It is applied  to the operators on \As\ and for any
operator $A$ on \As\ it satisfies
$ (f, A g)_\kappa = (A^{\ddagger} f, g)_\kappa $, as usual.
Since the class of hermitian realizations is only a subclass of the general
type of realizations that are considered in the paper, the star product
corresponding to any hermitian realization is also given by the expression (\ref{generalstarproducthahi}), with the
additional remark that $\chi$ has to be of a particular form to
provide for the hermiticity of $\hat{x}$. 

 Let us consider the following Fourier decompositions
\begin{equation} \label{comfourierexpansion}
  \phi (x) = \int \frac{\mathrm{d}^n k}{{(2\pi)}^{ n}} ~ \tilde{\phi} (k) ~ e^{ikx},
\end{equation}
for a generic element of the commutative algebra \An\ and
\begin{equation} \label{noncomfourierexpansion}
  \hat{\phi} (\hat{x}) = \int \frac{\mathrm{d}^n k}{{(2\pi)}^{ n}}
  ~ \tilde{\phi} (k) ~ \hat{e}_k^+ (\hat{x}),
\end{equation}
for the general element of \Ah\ .
Then the generalized complex conjugation acts according to
\begin{equation} \label{comfourierexpansiongenconj}
  {\phi}^* (x) = \int \frac{\mathrm{d}^n k}{{(2\pi)}^{ n}} ~
       \overline{\left(\tilde{\phi} (k)
    \right)} ~ {(e^{ikx})}^*,  
\end{equation}
for $f\in\As$.

Having this, it is now straightforward to establish what object does
the following map correspond to
\begin{eqnarray} \label{noncomfourierexpansiondagger}
  {\hat{\phi}}^{\dagger} (\hat{x}) \rhd 1 &=& \int \frac{\mathrm{d}^n k}{{(2\pi)}^{ n}}
  ~ \overline{ \tilde{\phi} (k)} ~ {(\hat{e}_k^+ (\hat{x}))}^{\dagger }
  \rhd 1  \nonumber\\
  &=& \int \frac{\mathrm{d}^n k}{{(2\pi)}^{ n}}
  ~ \overline{ \tilde{\phi} (k)} ~ \frac{ \mC_{h,\chi
 }({\mK}_h^{-1}(S_h (k)))}{\mC_{h,\chi }({\mK}_h^{-1}(k))}  \hat{e}_k^- (\hat{x})
  \rhd 1.
\end{eqnarray}
Due to relation (\ref{ekaminus}), this can be written as
\begin{equation} \label{noncomfourierexpansiondagger1}
  {\hat{\phi}}^{\dagger} (\hat{x}) \rhd 1  
  = \int \frac{\mathrm{d}^n k}{{(2\pi)}^{ n}}
  ~ \overline{ \tilde{\phi} (k)} 
  ~ \frac{ \mC_{h,\chi}({\mK}_h^{-1}(S_h (k)))}
          {\mC_{h,\chi }({\mK}_h^{-1}(k))} e^{i S_h(k) x}.
\end{equation}
Since we require the isomorphism (\ref{isomorphism})
 to be realized at the level of adjoint operators as well, 
that is ${\hat{\phi}}^{\dagger} (\hat{x}) \rhd 1 =
{\phi}^* (x)$, a direct comparison of this requirement with the
 relation (\ref{noncomfourierexpansiondagger1}) gives
 \begin{equation} \label{enaikaiksdager}
  {(e^{ikx} )}^{*} =  ~ \frac{ \mC_{h,\chi
  }({\mK}_h^{-1}(S_h (k)))}{\mC_{h,\chi }({\mK}_h^{-1}(k))} e^{i S_h
  (k) x}.
\end{equation}
Note that there is a class of star products  where the prefactor
appearing in (\ref{enaikaiksdager})
reduces to 1. One example of the star product with such a property is
the hermitian star product corresponding to the {\it{natural realization}}. 

With the correspondence just established, 
one can show the following identity involving the scalar product on \As:
\begin{equation}
  \int\mathrm{d}^nx\hat{f}^\dagger (\hat{x})\hat{g}(\hat{x})\rhd1=
    \int\mathrm{d}^nx f^*(x)\star g(x).
\end{equation}
The last expression can be more explicitly calculated 
for different types of star products on 
$\kappa$-Minkowski spacetime \cite{mssg,msa11}. For a
certain subclass of star products it can be shown to reduce to
$\ds \int\mathrm{d}^nx\overline{f(x)}g(x)$.


For a fixed $h$, and for any polynomial 
$P= P(\partial_0,\ldots,\partial_{n-1})$, we
define $S P = S_h(-i\partial_1,\ldots,-i\partial_{n-1})$.
Then for $f(x)=e^{ikx}$ and $g(x)=e^{iqx}$ one checks that 
\begin{equation}
  \int\mathrm{d}^nx f(x)\star\left(\pmu g(x)\right)=
    \int\mathrm{d}^nx\left((S\pmu)f(x)\star g(x)\right).
\end{equation}
By Fourier expansion this identity extends to general  
functions $f(x)$ and $g(x)$. The formula further generalizes 
to any polynomial $P$ in \pmu-s in the place of $\pmu$.

We note an interesting observation that
\begin{equation}
  g^\ddagger=S(\overline{g}),
\end{equation}
where $g$ is any element of the universal enveloping algebra
$U(\so(1,n-1)\rtimes\R^n)$ and bar denotes the
ordinary complex conjugation.
 
We conclude the set of identities with the formula 
\begin{equation}
  \int\mathrm{d}^nxf(x)\star g(x)=\int\mathrm{d}^nx\left(Z^\alpha g(x)\right)
    \star\left(Z^{-n+1+\alpha} g(x)\right),
\end{equation}
involving also the shift operator $Z$, which holds for
the star product on the $\kappa$-Minkowski spacetime
related to any of our realizations~(\ref{c1}).
It can easily be proved for the Fourier modes, i.e. for 
$f(x)=e^{ikx}$ and $g(x)=e^{iqx}$ and extending by bilinearity.

\section{General hermitian realization}
\label{ghr}

There is a procedure which from a given realization produces
many new realizations, using unitary similarity transformations.
Let
\begin{eqnarray}
  P_\mu&=&U^\dagger p_\mu U, \nonumber\\
  X_\mu&=&U^\dagger\xmu U, \label{e12}
\end{eqnarray}
for the unitary operator $U$ of the form
\begin{equation}
  U=\exp\{ i(x_\alpha\Sigma_\alpha+\tilde{\Sigma}-\Sigma_\alpha^\dagger x_\alpha
    -\tilde{\Sigma}^\dagger )\},
  \label{e13}
\end{equation}
where $\Sigma_\alpha$ and $\tilde{\Sigma}$ have the form
\begin{equation}
  \Sigma_\alpha=\Sigma_\alpha(A,B)=ip_{\alpha}\sigma_1(A,B)+ia_\alpha p^2
  \sigma_2(A,B)
\end{equation}
and $\tilde{\Sigma}=\sigma_3(A,B)$ for some functions $\sigma_i(A,B),
\; i = 1,2,3$. If one starts with a hermitian realization and
performs a similarity transformation according to Eq.(\ref{e12}), 
then this leads to another realization, which is also hermitian
under the assumption that $U$ is a unitary operator.
Thus, we demand that $\Sigma_\alpha$ in (\ref{e13}) 
are hermitian operators ($\sigma_1$ and $\sigma_2$ are antihermitian).
 In that case, $U$ tends to $I$ as $a$ tends to 0. Now
\begin{eqnarray}\label{e14}
  P_\mu&=&\exp\{-[x_\alpha\Sigma_\alpha+\tilde{\Sigma}-\Sigma_\alpha^\dagger
    x_\alpha-\tilde{\Sigma}^\dagger]\}p_\mu\exp\{x_\alpha\Sigma_\alpha+\tilde
    {\Sigma}-\Sigma_\alpha^\dagger x_\alpha-\tilde{\Sigma}^\dagger\}=\nonumber\\
  &=&p_\mu+\left(\Sigma_\mu-\Sigma_\mu^\dagger\right)+\frac1{2!}\left(
    \Sigma_\alpha-\Sigma_\alpha^\dagger\right)\frac{\partial}{\partial p_
    \alpha}\left(\sigma_\mu-\Sigma_\mu^\dagger\right)^\dagger +
    \ldots = \nonumber\\
  &=&p_\mu+\left(\frac{\exp O-1}O\right)
    \left(\Sigma_\mu-\Sigma_\mu^\dagger\right),
\end{eqnarray}
where
\begin{equation}
  O=\left(\Sigma_\alpha-\Sigma_\alpha^\dagger\right)\left(\frac{\partial}{\partial
   p_\alpha}\right).
\end{equation}

Suppose we want to find all hermitian realizations that are obtained
by means of the unitary transformations (\ref{e12}) from a given
hermitian realization, say $\hat{x}_{\mu}^{(\frac{1}{2})}$, characterized by $h_{\mu \nu}$.
Then, starting from $\hat{x}_{\mu}^{(\frac{1}{2})}$ for a fixed
 $h_{\mu \nu}$, one can generate
an infinite family of hermitian realizations with the same
   $h_{\mu \nu}$, but different $\chi_{\mu}$.
The condition on the unitary transformations $U$
 that perform this map and give a set of hermitian realizations with
 the same $h_{\mu \nu}$  is $\Sigma_\alpha = \Sigma^{\dagger}_\alpha = 0 $. 
Hence, it is possible to introduce 
   the set of unitary transformations~(\ref{e13}) that, after
   performing~(\ref{e12}), do not
   change $h_{\mu \nu}$ in~(\ref{c1}), but change only $\chi_\mu$; 
   those transformations are 
   specified by $\Sigma_\alpha = \Sigma^{\dagger}_\alpha = 0 $.

\section{Examples}

In this Section we apply the general results developed so far
by working them out on the four characteristic and important examples 
that realize the $\kappa$-Minkowski spacetime. Four instances in
question are the hermitizations of the following: left-covariant,
right-covariant, Weyl totally symmetric and {\it{natural}}
realization. 

 It is of interest to see how the star product
looks like in each of the four realizations mentioned. Crucial to this is
 to know the form of the $\mC(k,q)$ factors in each of
them. The knowledge of these factors also makes possible to calculate
the integral
$ \int \mathrm{d}^nx ~ e^{iS_h (k)x} ~ {\star}_{h,\chi } ~  e^{iqx}$ for every
and each of the  realizations. We particularly consider the hermitian
realizations corresponding to $\beta =\frac12 $ (that is, hermitian realizations obtained
through the standard symmetrization, $\hat{x}^{(\frac{1}{2})} = \frac{1}{2}
(\hat{x}_\mu + \hat{x}^{\dagger}_\mu ) $  (see Section 4)). 

 Before going to special cases, we restate the general results for the
 star product and a scalar product corresponding to the realization $h, \chi:$
\begin{equation}
  e^{ikx} ~ {\star}_{h,\chi } ~  e^{iqx} =
 \frac{\mC_{h,\chi } (\mK_h^{-1}(\mD_h (k, q) ))   }{\mC_{h,\chi }({\mK}_h^{-1}(k)) \mC_{h,\chi }({\mK}_h^{-1}(q)) }  
  e^{i \mD_h (k, q) x}. 
\end{equation}
This can also be expressed as

\begin{equation}
  e^{ikx} ~ {\star}_{h,\chi } ~  e^{iqx} =
 \frac{\mC_{h,\chi } (\mK_h^{-1}(k), q) }{\mC_{h,\chi
 }({\mK}_h^{-1}(k))}  
  e^{i \mD_h (k, q) x}. 
\end{equation}
The scalar product between two plane waves, with momenta $k$ and $q$ 
respectively, is thus proportional to
\begin{eqnarray}
{(e^{ikx}, e^{iqx} )}_{\kappa} & \sim & \int \mathrm{d}^nx ~  e^{iS_h (k)x} ~
  {\star}_{h,\chi } ~  e^{iqx} \nonumber \\
  &=& {(2\pi)}^n \frac{\mC_{h,\chi } (\mK_h^{-1}(\mD_h (S_h (k), q) ))
  }{\mC_{h,\chi }({\mK}_h^{-1}(S_h (k))) \mC_{h,\chi
  }({\mK}_h^{-1}(q)) } \frac{\delta^{(n)}(k-q)}{{\left|det\left(\frac{\partial(\mD_h (S_h(k),q))_\mu}
  {\partial q_\nu}\right)_{q=k}\right|}}   \nonumber \\
  &=&  {(2\pi)}^n \frac{\mC_{h,\chi } (\mK_h^{-1}(S_h (k)), q)    }{\mC_{h,\chi
 }({\mK}_h^{-1}(S_h (k)))}  
     \frac{\delta^{(n)}(k-q)}{{\left|det\left(\frac{\partial(\mD_h (S_h(k),q))_\mu}
  {\partial q_\nu}\right)_{q=k}\right|}}. \label{example} 
\end{eqnarray}

\subsection{Hermitization of the 
left covariant realization}

For the left covariant realization, $\ds \hat{x}_\mu=x_\mu(1-A) =
x_\mu (1 + ap)$, which means that
\begin{equation}
  h_{\mu \alpha} (p) = (1+ap) \eta_{\mu \alpha }
\end{equation}
and in this case  one gets 
\begin{equation}
  \mP_{h, \mu}(k,q) = k_\mu \frac{e^{ak} -1}{ak} (1+ aq) + q_\mu,
  \quad \mK_{h, \mu}(k) = k_\mu \frac{e^{ak} -1}{ak}.
\end{equation}
Consequently, $\mK^{-1}_{h, \mu}(k) = k_\mu \frac{\ln (1+ak)}{ak}$ and
\begin{equation}
  \mD_{h, \mu}(k,q) = \mP_{h, \mu}(\mK^{-1}_{h, \mu}(k),q) = k_\mu
  Z^{-1} (q) + q_\mu,
\end{equation}
where $Z^{-1}(q)=1+aq$.
The antipode in turn can be calculated by setting $k=S_h(q)$ and $\mD_{h,\mu} (k,q)=0$
in the previous expression:
\begin{equation}
  S_h(q)_\mu Z^{-1}(q)+q_\mu=0\;\Leftrightarrow S_h(q)_\mu = -q_\mu Z(q).
\end{equation}
Since $Z^{-1}(q)=1+aq$ for our realization,
\begin{equation}
  Z(q)=\frac1{1+aq}.
\end{equation}
Next we calculate the determinant appearing in (\ref{example}) for this particular
case of left covariant realization. In order to simplify calculations,
it is convenient to take $a=(a_0,0)$ and then, due to covariance,
to generalize the result to a four-vector $a_\mu$ oriented in an arbitrary direction,
\begin{eqnarray}
  &J(k)=\left|det \left(\frac{\partial\mD_{h,\mu}(S_h(k),q)}{\partial q_\nu}\right)_{q=k}\right|
  =\left|det \left(\frac{\partial\left(S_h(k)_\mu Z^{-1}(q)+q_\mu\right)}
  {\partial q_\nu}\right)_{q=k}\right|=&\nonumber \\
  &=\left|det\left|\left(S_h(k)_\mu a_\nu +\eta_{\mu\nu}\right)\right|_{q=k}\right|.&
\end{eqnarray}
This leads to
\begin{equation} \label{Jacobian}
  \det {\left( \frac{\partial {\mathcal{D}}_{h, \mu}(S_h(k),q)}{\partial q^{\lambda }} \right)}_{q=k} =
\left| \begin{array}{ccccc}
  \frac{1}{1 + ak} & 0  & 0 &
   \cdot \cdot \cdot 
     & ~  0 \\
 \frac{k_1 }{1 + ak}a_0   & 1 & 0 &  \cdot \cdot \cdot
 & 0  \\ 
 \frac{k_2 }{1 + ak} a_0  & 0 & 1 & \cdot \cdot \cdot 
 & 0  \\
 \cdot & \cdot &   \cdot \cdot \cdot & \cdot & \cdot \\
\cdot & \cdot &  \cdot \cdot \cdot & \cdot & \cdot  \\
  \frac{k_{n-1}}{1 + ak} a_0  & 0 & 0 & \cdot \cdot \cdot 
  & 1 \\
\end{array} \right| = \frac{1}{1+ak}.
\end{equation}
Following the hermitization procedure for the value $\beta =\frac12$, 
it is easy to get $\ds \hat{x}_\mu^{(\frac12)}=\hat{x}_\mu+\frac
i2a_\mu$ and thus read out the quantity $\chi_{\mu} (p) = \frac{i}{2}
a_\mu$, which is important for determining the $\mC(k,q)$ factors in
Eq. (\ref{example}). Due to $\chi$ being constant, the integration in (\ref{codkaku})
becomes trivial, giving $\mC_{h,\chi } (k,q) = \mC_{h,\chi } (k) =
e^{-\frac{1}{2} ak}$ and consequently
\begin{equation}  \label{resultsforleftcov}
 \mC_{h,\chi } (\mK_h^{-1} (k)) = \frac{1}{\sqrt{1+ak}},   \quad
 \mC_{h,\chi } (\mK_h^{-1}(S_h (k)), q) = \sqrt{1+ak}.
\end{equation}
With these results we have for the hermitized ($\beta =\frac12$) left
covariant realization
\begin{equation}
  e^{ikx} ~ {\star}_{h,\chi } ~  e^{iqx} = 
  e^{i \mD_h (k, q) x} 
\end{equation}
and the scalar product between two plane waves, with momenta $k$ and $q$, respectively
is thus proportional to
\begin{eqnarray}
{(e^{ikx}, e^{iqx} )}_{\kappa} & \sim & \int \mathrm{d}^nx ~  e^{iS_h (k)x} ~
  {\star}_{h,\chi } ~  e^{iqx}
  =  {(2\pi)}^n (1+ak) \delta^{(n)}(k-q).
\end{eqnarray}

\subsection{Hermitization of 
the right covariant realization}

For the right covariant realization, $\ds \hat{x}_\mu=x_\mu - a_\mu
x_{\alpha} p_{\alpha} $, which means that
\begin{equation}
  h_{\mu \alpha} (p) =  \eta_{\mu \alpha } - a_{\alpha} p_{\mu}.
\end{equation}
A similar  analysis as before, when applied to this particular situation, gives
\begin{equation}
  \mD_{h, \mu}(k,q)  = k_\mu + Z(k) q_\mu,
\end{equation}
with $Z(k)=1-ak$ ( $Z^{-1}(k)= 1/ (1-ak)$).
The opposite momentum, or the antipode is readily extracted from the
function $\mD_{h}$ to be $S_h(k)_{\mu} = -k_{\mu} Z^{-1} (k)$. One can
easily check that $Z(S_h(k)) = Z(k).$
Similarly as before the determinant for the case of right covariant realization can be calculated as
\begin{equation}
  J(k)=\left|det \left(\frac{\partial\mD_{h,\mu}(S_h(k),q)}{\partial q_\nu}\right)_{q=k}\right|
    =  Z^{n} (S_h(k))  = Z^{n} (k).
\end{equation}
Hermitization procedure for $\beta = \frac12$ identifies the 
$\chi $ to be ${\chi}_{\mu} (p) = \frac{in}{2} a_\mu$. This leads to
$\mC_{h,\chi } (k,q) = \mC_{h,\chi } (k) =
e^{-\frac{n}{2} ak}$ and consequently the orthonormality property of the
plane waves in the right covariant realization can be expressed as
\begin{eqnarray}
{(e^{ikx}, e^{iqx} )}_{\kappa}  \sim  \int \mathrm{d}^nx ~  
e^{iS_h (k)x} ~ {\star}_{h,\chi } ~  e^{iqx}
  &=&  {(2\pi)}^n Z^{-n}(k) \delta^{(n)}(k-q)   \nonumber  \\
  &=& \frac{{(2\pi)}^n}{{(1-ak)}^{n}} \delta^{(n)}(k-q).
\end{eqnarray}

\subsection{Hermitization of the Weyl symmetric realization}
For the Weyl symmetric realization $\hat{x}_\mu = x_\mu h_s - a_\mu
xp \gamma_1$, that is
\begin{equation}
  h_{\mu \alpha} (p) =  \eta_{\mu \alpha } h_s - a_{\alpha} p_{\mu} \gamma_1,
\end{equation}
where
\begin{equation}
   h_s (A) = \frac{A}{e^A -1},  \quad  \gamma_1 (A) = \frac{1}{A} (1 -
   \frac{A}{e^A -1} ) \equiv \frac{1 - h_s (A)}{A}, \quad   A= -ak.
\end{equation}
By repeating the same steps as before, we get
\begin{equation}
  \mP_{h,\mu} (k,q) = \mD_{h,\mu} (k,q)=h_s(k+q)\left(\frac{k_\mu}{h_s(k)}+Z(k)
  \frac{q_\mu}{h_s(q)}\right),
\end{equation}
where $Z(k) = e^A = e^{-ak}$, $Z(S_h(q))=e^{-aS_h(k)}$.
Again, the antipode can be deduced by setting $k = S_h(q)$ and $\mD_{h,\mu}
(k,q) =0$,
\begin{equation}
  \mD_{h,\mu} (S_h(k),q)=h_s(S_h(k)+q)
\left(\frac{S_h(k_\mu)}{h_s(S_h(k))}+Z(S_h(k))
  \frac{q_\mu}{h_s(q)}\right).
\end{equation}
This implies that the antipode $S_h(q)$ has to satisfy the relation
\begin{equation}
  \frac{S_h(q_\mu)}{h_s(S_h(q))}+Z(S_h(q))\frac{q_\mu}{h_s(q)}=0.
\end{equation}
The solution is $S_h(k)_\mu = -k_\mu$, in consistency with 
the results in Appendix B. 
 The calculation of the determinant in (\ref{example}) gives
\begin{equation}
  J(k)=\left|det \left(\frac{\partial\mD_{h,\mu}(S_h(k),q)}{\partial q_\nu}\right)_{q=k}\right|
    =  {\left( \frac{Z^{-1}(k)}{h_s (k)}  \right)}^{n-1},  
\qquad  h_s (k) = \frac{ak}{1- e^{-ak}}
\end{equation}
where $h_s (k) = \frac{ak}{1- e^{-ak}}.$
Again, the hermitization procedure for $\beta =\frac12$ gives rise to
$\chi $, which is this time nontrivial:
 $$
{\chi}_{\mu} (p) = \frac{i}{2} a_\mu (\frac{\partial h_s (A)}{\partial A}
+ n \gamma_1 (A) - ap \frac{\partial \gamma_1 (A)}{\partial A} ) =
\frac{i}{2} a_\mu (n-1) \gamma_1 (A).
$$
 This leads to
$$
\mC_{h,\chi } (k,q) = {\left( \frac{ak+aq}{aq} \frac{(e^{-aq} -1) e^{-ak}}{e^{-(ak+aq)} -1}  \right)}^{\frac{n-1}{2}}
   = {\left( \frac{h_s (k+q)}{h_s (q)} Z(k)   \right)}^{\frac{n-1}{2}}
$$ 
 and consequently the orthonormality property for the
plane waves in the Weyl symmetric realization can be expressed as
\begin{eqnarray} \label{Weylsymmetricscalarproduct}
{(e^{ikx}, e^{iqx} )}_{\kappa}  \sim  \int \mathrm{d}^nx ~  e^{iS_h (k)x} ~ {\star}_{h,\chi } ~  e^{iqx}
  &=&  {(2\pi)}^n Z^{\frac{n-1}{2}}(k) \delta^{(n)}(k-q),  \quad  Z(k)
  = e^{-ak}. 
\end{eqnarray}
However, it is interesting to note that the ordinary plane waves are 
orthonormalized for the scalar product $(,)_\kappa$, in the case of
Weyl symmetric ordering,
\begin{equation}\label{orthonorm}
{(e^{ikx}, e^{iqx} )}_{\kappa}  ~
  = ~ {(2\pi)}^n  \delta^{(n)}(k-q).
\end{equation}
It is seen from  (\ref{starproductonastar}) and (\ref{enaikaiksdager})
that, in the case of the Weyl symmetric star product, 
the additional factor 
$\frac{ \mC_{h,\chi  }({\mK}_h^{-1}(S_h (k)))}{\mC_{h,\chi }({\mK}_h^{-1}(k))}$ 
in~(\ref{enaikaiksdager})
 cancels out the corresponding factor in
 (\ref{Weylsymmetricscalarproduct}), giving rise to the
 orthonormalization property~(\ref{orthonorm}). 
In the next subsection, we shall see
that the same property holds when the scalar product is introduced
in terms of the {\it{natural}} realization. Therefore, 
as it was the case with the Weyl symmetric realization, the plane
waves are also orthonormal in the {\it{natural}} realization.
This feature was noticed for the first time in~\cite{msa11}.
In contrast, for the left and right covariant realizations our
analysis shows that this property does not hold.

\subsection{Hermitization of the {\it{natural}} realization}

The analysis for the {\it{natural}} realization has already been
carried out in~\cite{msa11}, so here we only recapitulate the
results obtained there and compare them with other realizations. It is
specified by the following differential representation for the
generators of the $\kappa$-Minkowski algebra,
\begin{equation} \label{hn0}
 {\hat{x}}^{h}_{\mu}  =   x_{\mu}(a_{\alpha}{p}^{\alpha}+\sqrt{1+a^2{p}_{\alpha}{p}^{\alpha}})-(ax){p}_{\mu},
\end{equation}
or in another way, it is specified by
\begin{equation} \label{hn00}
  h_{\alpha \mu} (p) = {\eta}_{\alpha \mu}(ap + \sqrt{1 + a^2 p^2 })
  -a_\alpha p_\mu.
\end{equation}
This leads to $\mD_h (k,q)$ with the form
\begin{equation} \label{hn0a}
  \left(\mD_h(k,q)\right)_\mu=k_\mu Z^{-1}(q)+q_\mu-a_\mu(kq)Z(k)+
    \frac12a_\mu(aq)\square(k)Z(k),
\end{equation}
where
\begin{equation} \label{hn0b}
  Z^{-1} (k) ~ \equiv ~ ak + \sqrt{1 + a^2 k^2}, \qquad
   \square(k) ~ \equiv ~ \frac{2}{a^2}\bigg[ 1- \sqrt{1 + a^2 k^2}
   \bigg]. 
\end{equation}
The determinant in (\ref{example}) was found to be
\begin{equation} \label{hn4}
   J(k)= {\left|det\left(\frac{\partial(\mD_h (S_h(k),q))_\mu}
  {\partial q_\nu}\right)_{q=k}\right|} = \frac{1}{\sqrt{1+a^2k^2}},
\end{equation}
so that ${\delta}^{(n)} \left( {\mathcal{D}_h}(S_h(k),q)\right) ~ = ~
 \sqrt{1+ a^2 k^2} ~ {\delta}^{(n)} (q-k ).$
The hermitization procedure, similarly to the previous cases, produces $\chi$
congruent to the $\beta = \frac{1}{2}$ hermitization to be
\begin{equation} \label{hn01}
  \chi_{ \mu} (p) =  -\frac{i}{2} \frac{a^2 p_\mu}{ \sqrt{1 + a^2 p^2 }}.
\end{equation}
Using the formula for the additional factor $\mC_{h, \chi}$ given by
(\ref{codkaku}), in the case of {\it{natural}} realization we obtain
\begin{equation}
  \mC_{h,\chi }({\mK}_h^{-1}(k)) = \sqrt[4]{1+a^2k^2},
\end{equation}
\begin{equation}
  \mC_{h,\chi }({\mK}_h^{-1}(k), q) = \frac{\sqrt[4]{1+a^2  {\mD}_h^2 (k,q)}}{\sqrt[4]{1+a^2 q^2}}.
\end{equation}
Also, the star product (\ref{generalstarproducthahi}) transforms to
\begin{eqnarray}
  e^{ikX} ~ {\star}_{h,\chi } ~  e^{iqX}  &=&
 \frac{\mC_{h,\chi } (\mK_h^{-1}(\mD_h (k, q) ))   }{\mC_{h,\chi }({\mK}_h^{-1}(k)) \mC_{h,\chi }({\mK}_h^{-1}(q)) }  
  e^{i \mD_h (k, q) X} \label{hn1} \\
  &=&
 \frac{\sqrt[4]{1+a^2  {\mD}_h^2 (k,q)}}{{\sqrt[4]{1+a^2k^2}}{\sqrt[4]{1+a^2q^2}}}  
  e^{i \mD_h (k, q) X}, \label{hn5}
\end{eqnarray}
which in turn gives rise to
\begin{equation}
{(e^{ikx}, e^{iqx} )}_{\kappa}  =  \int \mathrm{d}^nx ~  e^{iS_h (k)x} ~
  {\star}_{h,\chi } ~  e^{iqx} 
  = {(2\pi)}^n \delta^{(n)}(k-q).\nonumber
 \end{equation}
\\

In summary we arrange the results in the tabular form

\begin{tabular}{|r||c|c|}\hline
                        & Left covariant & Right covariant \\ \hline\hline
 $h_{\mu \alpha} (k)$   &   $ (1+ak)\eta_{\mu \alpha} $          & $\eta_{\mu \alpha} - a_\alpha k_\mu $ \\ \hline
  $\chi_\mu (k)$        &    $\frac{i}{2}a_\mu $            & $\frac{in}{2}a_\mu $ \\ \hline
  $Z^{-1}(k)$               & $1+ak $            &       $\frac{1}{1-ak} $                                \\ \hline
  $J(k)$               & $\frac{1}{1+ak} $            &       $Z^{n}(k)$                                 \\ \hline
  $\mK^{-1}_{h,\mu }(k)$              &     $\frac{\ln(1+ak)}{ak} k_\mu$            & $-\frac{\ln(1-ak)}{ak} k_\mu$ \\ \hline
  $\mC_{h,\chi }(k,q)$    &     $e^{-\frac{1}{2} ak}$            &   $e^{-\frac{n}{2} ak}$        \\ \hline
  $\mC_{h,\chi }(k)$      &      $e^{-\frac{1}{2} ak}$           &  $e^{-\frac{n}{2} ak}$  \\ \hline
  $\mC_{h,\chi }(\mK^{-1}_{h }(S_h (k)),q)$&    $\sqrt{1+ak}$         &  $\frac{1}{{(1-ak)}^{\frac{n}{2}}}$ \\ \hline
  $\mC_{h,\chi }(\mK^{-1}_{h }(k))$&    $\frac{1}{\sqrt{1+ak}}$           &  ${(1-ak)}^{\frac{n}{2}}$ \\ \hline
  $ I(k)$   &    $1+ak$            & $\frac{1}{{(1-ak)}^{n}}$ \\ \hline
  $ {(e^{ikx}, e^{iqx} )}_{\kappa}$  &    $ {(2\pi)}^n {(1+ak)}^2 \delta^{(n)}(k-q)$ &  $ {(2\pi)}^n \frac{1}{{(1-ak)}^{2n}} \delta^{(n)}(k-q) $ \\ \hline
\end{tabular}

\begin{tabular}{|r||c|c|}\hline
                        & Weyl symmetric & Natural\\ \hline\hline
 $h_{\mu \alpha} (k)$   & $\eta_{\mu \alpha}h_s - a_\alpha k_\mu \gamma_1$  & ${\eta}_{\mu \alpha }(ak + \sqrt{1 + a^2 k^2 }) -a_\mu k_\alpha $\\ \hline
  $\chi_\mu (k)$        &    $\frac{i}{2} a_\mu (n-1) \gamma_1 $    & $-\frac{i}{2} \frac{a^2 k_\mu}{ \sqrt{1 + a^2 k^2 }}$ \\ \hline
  $Z^{-1}(k)$               &$ e^{ak}$            &  $ ak + \sqrt{1 + a^2 k^2} $                                  \\ \hline
  $J(k)$               &${\left( \frac{Z^{-1}(k)}{h_s (k)}  \right)}^{n-1}$      & $\frac{1}{\sqrt{1+a^2 k^2}}$   \\ \hline
  $\mK^{-1}_{h, \mu }(k)$  &     $k_\mu$ & $\bigg[ k_{\mu} - \frac{a_{\mu}}{2} \square(k)
       \bigg] \frac{\ln \left( Z^{-1} (k) \right)}{ Z^{-1} (k) -1}$  \\ \hline
  $\mC_{h,\chi }(k,q)$  &  ${\left( \frac{h_s (k+q)}{h_s (q)} Z(k)
 \right)}^{\frac{n-1}{2}}$  & $\frac{\sqrt[4]{1+a^2  {\mD}_h^2 (\mK_h
 (k),q)}}{\sqrt[4]{1+a^2 q^2}}$ \\ \hline
  $\mC_{h,\chi }(k)$      &  ${\left( h_s (k) Z(k)
 \right)}^{\frac{n-1}{2}}$  & $\sqrt[4]{1+a^2 {(\mK_h (k))}^2}$  \\ \hline
  $\mC_{h,\chi }(\mK^{-1}_{h }(S_h (k)),q)$&  ${\left( \frac{h_s (-k+q)}{h_s (q)} Z(-k)
 \right)}^{\frac{n-1}{2}}$  & $\frac{\sqrt[4]{1+a^2  {\mD}_h^2 (S_h
 (k),q)}}{\sqrt[4]{1+a^2 q^2}}$ \\ \hline
  $\mC_{h,\chi }(\mK^{-1}_{h }(k))$& ${\left( h_s (k) Z(k) \right)}^{\frac{n-1}{2}}$          &  $\sqrt[4]{1+a^2k^2}$ \\ \hline
  $ I(k)$   &   $ Z^{\frac{n-1}{2}}(k) $          & $1$ \\ \hline
  $ {(e^{ikx}, e^{iqx} )}_{\kappa}$  & $ {(2\pi)}^n  \delta^{(n)}(k-q) $             & ${(2\pi)}^n  \delta^{(n)}(k-q)$ \\ \hline
\end{tabular}
\\
In the above table $h_s = \frac{ak}{1- e^{-ak}}$ and $\gamma_1 = \frac{h_s -1}{ak}$.
\\

It has to be emphasized that in all four cases presented in the
tables, the function $\chi_\mu (k)$ is chosen so that it corresponds
to the hermitization procedure determined by the parameter $\beta =
\frac{1}{2}$ (see Section 4). On the other side, the scalar product
(the last line in the tables) is calculated by using
Eqs.(\ref{starproductonastar}) and (\ref{enaikaiksdager}) as
\begin{eqnarray} 
 {(e^{ikx}, e^{iqx} )}_{\kappa}  &=& \frac{ \mC_{h,\chi
  }({\mK}_h^{-1}(S_h (k)))}{\mC_{h,\chi }({\mK}_h^{-1}(k))}  ~ \int
  \mathrm{d}^nx ~  e^{iS_h (k)x} ~ {\star}_{h,\chi } ~  e^{iqx}, \nonumber \\
\int \mathrm{d}^nx ~  e^{iS_h (k)x} ~ {\star}_{h,\chi } ~  e^{iqx}
   & =&  {(2\pi)}^n I(k) \delta^{(n)}(k-q).   \nonumber \\
\end{eqnarray}
From the table it is readily seen that the plane waves with different
momenta constitute an orthonormal set of vectors in Weyl
symmetric and {\it{natural}} basis, while in left covariant and right
covariant do not.
 This property is rather appealing because it implies that upon integration the star product of two functions can be replaced by an ordinary multiplication. It is still an open question whether this property can be related in some way to the trace property of the action integral or at least shed some light on this otherwise a notoriously difficult problem. The same constatation applies to the problem of finding the appropriate integration measure in the action integral valid for $\kappa$-type of deformation, which is a crucial ingredient in discussing the noncommutative action integral defined on the $\kappa$-deformed space.
In turn both of these issues are intimately related to the proper introduction of a gauge theory in the above setting. Namely, when introducing
a gauge field into the theory, then one might have the trace property satisfied by the
gauge field integral because it ensures a gauge invariance of the action and a theory.
We point out that in general  different realizations of 
$\kappa$-Minkowski spacetime might have different physical consequences (see 
also~\cite{Meljanac:2012pv,gpmp09}).

\section*{Acknowledgements}

We are very thankful to Andrzej Borowiec for the numerous comments on the text.
Discussions with Jerzy Lukierski were very valuable.
Also, Tajron Juri\cm\ is acknowledged for remarks.
A.S. thanks the Galileo Galilei Institute for Theoretical Physics (Florence)
for the hospitality and the INFN for partial support during which a part
of this work was done.
This work was also supported by the Ministry of Science, Education and Sports 
of the Republic of Croatia under contract No. 098-0000000-2865.

\appendix
\numberwithin{equation}{section}

\section{Calculation of $\mC(k,q)$ }

  In this section we prove the relation (\ref{codkaku}). Let us start with
  the defining  relation for $\mC_{h,\chi }(k,q)$ and $\mP_{h}(k,q)$,
\begin{equation}    \label{a1}
  e^{ik\hat{x}_{h,\chi }}\rhd e^{iqx}=\mC_{h, \chi}(k,q)e^{i\mP_{h}(k,q)x},
\end{equation}
where ${(\hat{x}_{h,\chi })}_{\mu} = {\hat{x}}_{\mu} + {\chi}_{\mu} (p) $, with
 ${\hat{x}}_{\mu} = x_{\alpha} h_{\alpha \mu} (p) $. The procedure
below determines $\mC_{h,\chi}$ and $\mP_{h}$ showing existence and 
uniqueness,hence that the ansatz~(\ref{a1}) indeed defines them. 
Alternatively, the form~(\ref{a1}) can be justified 
by generalizing the reasoning 
with power series in~\cite{meljSkodaSvrtan}, Section 2. 
For $\chi = 0$, ${(\hat{x}_{h,\chi })}_{\mu} = {\hat{x}}_{\mu}$
and $e^{ik\hat{x}}\rhd e^{iqx}=e^{i\mP_{h}(k,q)x}$, implying that
 $\mC_{h,\chi = 0 }(k,q) =1$.  If we define the
 family of operators  
\begin{equation} \label{a2}
  P_{\mu}^{(t)}(k,-i \partial ) ~ = ~ e^{-itk \hat{x}} (-i {\partial}_{\mu}) e^{itk \hat{x}}, \qquad 0\leq t\leq 1,
\end{equation}
parametrized by the free parameter $t$, then it can be shown (see~\cite{mmss,mmssM}) that they
satisfy the differential equation
\begin{equation} \label{a3}
 \frac{dP_{\mu}^{(t)}(k,-i \partial)}{dt} ~ = ~
 h_{\mu\alpha}(P^{(t)}(k,-i \partial)) k_{\alpha},
\end{equation}
and that they are related to $\mP_{h}(k,q)$ as $\mP_{h, \mu}(k,-i\partial) = P_{\mu}^{(1)}(k,-i\partial). $
Hence, the link of the  operator family~(\ref{a2}) 
with $\mP_{h}(k,q)$ is obvious after making
the identification $q=-i\partial $. One just needs to take care about the
boundary condition for the solution of the
differential equation (\ref{a3}), which can be put in the form
 $P_{\mu}^{(0)}(k,-i\partial) = -i\partial_{\mu} \equiv q_{\mu}.$
Note also that $P_{\mu}^{(t)}(k,q) =P_{\mu}^{(1)}(tk,q) = \mP_{h, \mu}(tk,q) $.

To determine the factor $\mC_{h,\chi }(k,q)$, act from the left on
both sides of the Eq.(\ref{a1}), first  with the operator
${\partial}_{\mu}$ and then with the operator $e^{-ik\hat{x}}$ to yield 
\begin{equation} \label{a4}
  e^{-ik \hat{x}}  {\partial}_{\mu} e^{ik {\hat{x}}_{h, \chi}}
 = \mC_{h, \chi}(k,q) i \mP_{h, \mu}(k,q).
\end{equation}
Make the exchange $k \rightarrow tk$ in (\ref{a4})
and then differentiate both sides of the resulting expression with
respect to $t$ to get
\begin{eqnarray} \label{a5}
 &  e^{-itk \hat{x}} \bigg[i k_{\alpha} [{\partial}_{\mu},
 {\hat{x}}_{\alpha}] + {\partial}_{\mu} i k_{\alpha} {\chi}_{\alpha} (p)
 \bigg] e^{itk {\hat{x}}_{h, \chi}} \nonumber  \\
 & = i \frac{d\mC_{h, \chi}(tk,q)}{dt} \mP_{h, \mu}(tk,q) +i\mC_{h,
 \chi}(tk,q) \frac{d\mP_{h, \mu}(tk,q)}{dt}.
\end{eqnarray}
Due to $[p_{\mu}, {\hat{x}}_{\alpha}] = [p_{\mu}, {(\hat{x}_{h,\chi
  })}_{\alpha}] = -i h_{\mu \alpha} (p)$, the last expression turns into
\begin{eqnarray} \label{a6}
 & i k_{\alpha} \bigg[ e^{-itk \hat{x}} h_{\mu \alpha} (p) e^{itk {\hat{x}}_{h, \chi}}
  + e^{-itk \hat{x}} {\partial}_{\mu} {\chi}_{\alpha} (p)
  e^{itk {\hat{x}}_{h, \chi}}  \bigg] \nonumber  \\
 & = i \frac{d\mC_{h, \chi}(tk,q)}{dt} \mP_{h, \mu}(tk,q) +i\mC_{h,
 \chi}(tk,q) \frac{d\mP_{h, \mu}(tk,q)}{dt}.
\end{eqnarray}
It still remains to calculate each of the terms in square brackets. To obtain
the first one, we proceed by applying the operator $h_{\mu \alpha} (p)$
to both sides of Eq.(\ref{a1}) from the left (note that the argument
$p$ in $h_{\mu \alpha} (p)$ is an operator )
\begin{eqnarray}    \label{a7}
 h_{\mu \alpha} (p) e^{itk\hat{x}_{h,\chi }}&\rhd\,\,e^{iqx} = \mC_{h,
 \chi}(tk,q) h_{\mu \alpha} (-i \partial) e^{i\mP_{h}(tk,q)x}, \nonumber \\
 &= \mC_{h, \chi}(tk,q) h_{\mu \alpha} (\mP_{h}(tk,q)) e^{i\mP_{h}(tk,q)x}
\end{eqnarray}
which, after multiplying by $ e^{-itk \hat{x}},$ gives
\begin{equation}    \label{a8}
  e^{-itk \hat{x}} h_{\mu \alpha} (p) e^{itk\hat{x}_{h,\chi }}\rhd e^{iqx}=\mC_{h,
 \chi}(tk,q) h_{\mu \alpha} (\mP_{h}(tk,q) ) e^{-itk \hat{x}} \rhd  e^{i\mP_{h}(tk,q)x}.
\end{equation}
Since $e^{-itk \hat{x}} \rhd  e^{i\mP_{h}(tk,q)x} = e^{iqx}$, we finally have
\begin{equation}    \label{a9}
  e^{-itk \hat{x}} h_{\mu \alpha} (p) e^{itk\hat{x}_{h,\chi }}  =
\mC_{h, \chi}(tk,q) h_{\mu \alpha} (\mP_{h}(tk,q) ).
\end{equation}
Analogously, for the second term in (\ref{a6}) one gets
\begin{equation}    \label{a10}
  e^{-itk \hat{x}} \partial_{\mu} \chi_{ \alpha} (-i\partial) e^{itk\hat{x}_{h,\chi }}  =
 i \mC_{h, \chi}(tk,q)      \chi_{ \alpha} (\mP_{h}(tk,q) ) \mP_{h, \mu}(tk,q).
\end{equation}
Plugging last two expressions back to Eq.(\ref{a6}) gives
\begin{eqnarray} \label{a11}
 & i k_{\alpha} \bigg[ \mC_{h, \chi}(tk,q) h_{\mu \alpha} (\mP_{h}(tk,q) )
  + i \mC_{h, \chi}(tk,q)      \chi_{ \alpha} (\mP_{h}(tk,q) ) \mP_{h, \mu}(tk,q)  \bigg] \nonumber  \\
 & = i \frac{d\mC_{h, \chi}(tk,q)}{dt} \mP_{h, \mu}(tk,q) +i\mC_{h,
 \chi}(tk,q) \frac{d\mP_{h, \mu}(tk,q)}{dt}.
\end{eqnarray}
Due to the differential equation (\ref{a3}), 
the first term on the left-hand side and
the second term on the right-hand side of the above relation cancel each other,
producing the following differential equation
\begin{equation}    \label{a12}
  \frac{d\mC_{h, \chi}(tk,q)}{dt} = i \mC_{h, \chi}(tk,q) k_{\alpha} \chi_{ \alpha} (\mP_{h}(tk,q) ).
\end{equation}
This finally integrates to desired result, Eq.(\ref{codkaku}), which we have set to prove.

\section{Useful identities}

  We prove the following identity:
\begin{equation} \label{importantidentity}
  {\mK}_h ( \mD_{s } ( {\mK}_h^{-1}(k), {\mK}_h^{-1}(q))) = \mD_{h }(k,q). 
\end{equation}
It should first be noted that for each realization 
there is an ordering prescription corresponding to
it. Thus, for the realization specified by the functions $h$ and $
\chi$, Eq.~(\ref{c1}), the corresponding ordering we designate by the symbol  $:~:_{h, \chi}$. This correspondence is realized through the requirement
\begin{equation} \label{realizationorderingcorrespondence}
  : e^{ik \hat{x}_{h, \chi}} :_{h, \chi} \rhd 1 = \mC_{h, \chi} (K_{h, \chi} (k)) e^{ikx}
\end{equation}
where $\mC_{h, \chi}$ is defined in relation (\ref{definitionofk}) and  the function $K_{h, \chi} (k)$ specifies the ordering prescription
in such a way that
\begin{equation}
 : e^{ik \hat{x}} :_{h, \chi} =  e^{iK_{h, \chi} (k)  \hat{x}}. 
\end{equation}
Under  the substitution $p = K_{h, \chi} (k) $ and relabeling $p \rightarrow k $
 relation (\ref{realizationorderingcorrespondence}) takes the form
\begin{equation} 
   e^{ik \hat{x}_{h, \chi}}  \rhd 1 = \mC_{h, \chi} (k) e^{i K^{-1}_{h, \chi} (k) x}.
\end{equation}
A direct comparison of this with (\ref{definitionofk}) gives the relationship
\begin{equation}  \label{relationshipbetweenkandk}
   \mK_{h} (k) = K^{-1}_{h, \chi} (k),
\end{equation}
showing that $K_{h, \chi} (k) \equiv K_{h} (k)  $ also does not depend on $\chi $.

The definition of the star product (\ref{starproductdefinition}) makes
possible to calculate $e^{ikx} {\star}_{h, \chi} e^{iqx} $, but this
time with the help of the isomorphism (\ref{isomorphism}) specified by
the correspondence (\ref{realizationorderingcorrespondence}),
\begin{equation}\begin{array}{lcl}
   e^{ikx} ~ {\star}_{h, \chi}~ e^{iqx} &=&
   \frac{1}{\mC_{h, \chi} (K_{h, \chi} (k))} : e^{ik \hat{x}_{h, \chi}} :_{h, \chi}
  \frac{1}{\mC_{h, \chi} (K_{h, \chi} (q))} : e^{iq \hat{x}_{h, \chi}}
   :_{h, \chi} \rhd ~ 1 \\
  &=& \frac{1}{\mC_{h, \chi} (K_{h, \chi} (k)) 
\mC_{h, \chi} (K_{h, \chi} (q)) }   
   : e^{i(k ~ {\oplus} _{h, \chi} ~ q) 
\hat{x}_{h, \chi}} :_{h, \chi} \rhd ~ 1 \\
   &=& \frac{1}{\mC_{h, \chi} (K_{h, \chi} (k)) \mC_{h, \chi} (K_{h, \chi} (q)) }   
   : e^{i \mD_h (k,q) \hat{x}_{h, \chi}} :_{h, \chi} \rhd ~ 1.
\end{array}\end{equation}
Finally, due to (\ref{realizationorderingcorrespondence}), the last
expression transforms into
\begin{equation} \label{starprodizracunatnadruginacin}
   e^{ikx} ~ {\star}_{h, \chi} ~ e^{iqx}  =
    \frac{\mC_{h, \chi} (K_{h, \chi} (\mD_h (k,q)))}{\mC_{h, \chi} (K_{h, \chi} (k)) \mC_{h, \chi} (K_{h, \chi} (q)) }   
    e^{i \mD_h (k,q) x}.
\end{equation}
This is the same star product as that given in
Eq.(\ref{generalstarproducthahi1}), but only calculated in a different way.
Hence, a direct comparison of these two, together with (\ref{relationshipbetweenkandk}), gives directly the relation (\ref{importantidentity}).

 The identity (\ref{importantidentity}) can be specialized to $k
 \rightarrow S_h(k) $ to yield
\begin{equation} \label{josjedanvazanidentitet}
   e^{i \mD_h (S_h (k),q) x}  =
   e^{i {\mK}_h ( \mD_{s } ( {\mK}_h^{-1}(S_h (k)), {\mK}_h^{-1}(q)))  x}.
\end{equation}
By letting $k = q$ we have $\mD_h (S_h (k),k) = 0$, thus obtaining $1$
on the left-hand side of (\ref{josjedanvazanidentitet}). This means that the
right hand side also has to be equal to $1$, yielding
\begin{equation} \label{josjedanvazanidentiteta}
  {\mK}_h ( \mD_{s } ( {\mK}_h^{-1}(S_h (k)), {\mK}_h^{-1}(k))) = 0.
\end{equation}
Since ${\mK}_h^{-1}(0) = 0$ and $\mD_{s } (S_s (k), k) = 0$, relation (\ref{josjedanvazanidentiteta})
in turn implies
\begin{equation} \label{josjedanvazanidentitet1}
   {\mK}_h^{-1}(S_h (k)) = S_s ( {\mK}_h^{-1}(k)) = -{\mK}_h^{-1}(k),
\end{equation}
where in the final step the obvious property, $S_s (k) = -k$,
has been used. Here $S_s (k)$ is the antipode of $k$ corresponding to the
momentum addition rule in the Weyl totaly symmetric ordering,
$S_s (k) \oplus_s k \equiv \mD_{s } (S_s (k), k) = 0 $ in the same way
as $S_h (k)$, satisfying  $S_h (k) \oplus_{h, \chi} k \equiv \mD_{h } (S_h (k), k) = 0 $,
defines the antipode corresponding to a generic ordering prescription
$:~:_{h, \chi}$.

\section{Calculation of $\mC(k)$}

In order to calculate $\mC(k)$, we restrict to the set of realizations
whose corresponding star product satisfies
\begin{equation}\label{g1}
  \int d^n x e^{i S_h (k)x} ~\star_{h,\chi } ~ e^{iqx}= {(2\pi)}^n
  {\delta}^{(n)} (k-q).
\end{equation}
Now, the general expression 
 for the star product~(\ref{generalstarproducthahi}),
 applied for $k \rightarrow S_h (k)$,  shows that
\begin{equation} \label{g2}
 \int d^n x e^{i S_h (k)x} ~ {\star}_{h,\chi } ~  e^{iqx} = 
  \frac{\mC_{h,\chi } (\mK_h^{-1}(\mD_h (S_h (k), q) ))   }{\mC_{h,\chi }({\mK}_h^{-1}(S_h (k))) \mC_{h,\chi }({\mK}_h^{-1}(q)) }  
 \int d^n x  e^{i \mD_h (S_h (k), q) x}.
\end{equation}
The last expression gives a $\delta$-function which is readily calculated as
\begin{eqnarray} 
 \int d^n x e^{i S_h (k)x} ~ {\star}_{h,\chi } ~  e^{iqx} & =& 
  \frac{\mC_{h,\chi } (\mK_h^{-1}(\mD_h (S_h (k), q) ))   }{\mC_{h,\chi }({\mK}_h^{-1}(S_h (k))) \mC_{h,\chi }({\mK}_h^{-1}(q)) }  
  {(2\pi)}^n {\delta}^{(n)} (\mD_h (S_h (k), q)) \nonumber \\
 &=& {(2\pi)}^n \frac{\mC_{h,\chi } (\mK_h^{-1}(\mD_h (S_h (k), q) ))   }{\mC_{h,\chi }({\mK}_h^{-1}(S_h (k))) \mC_{h,\chi }({\mK}_h^{-1}(q)) }   
   \frac{{\delta}^{(n)} (k-q)}{\left|det\left(\frac{\partial(\mD_h (S_h(k),q))_\mu}
  {\partial q_\nu}\right)\right|}. \nonumber
\end{eqnarray}
Since this is equal to ${(2\pi)}^n {\delta}^{(n)} (k-q)$, we get a
condition
\begin{equation} \label{g3}
  \mC_{h,\chi }({\mK}_h^{-1}(S_h (k))) \mC_{h,\chi }({\mK}_h^{-1}(k))   
   = \frac{1}{\left|det\left(\frac{\partial(\mD_h (S_h(k),q))_\mu}
  {\partial q_\nu}\right)_{k=q}\right|},
\end{equation}
 after setting $k=q$. This is because $\mD_h (S_h(k),k)=0$, and
 consequently $ \mC_{h,\chi } (\mK_h^{-1}(\mD_h (S_h (k), k) ))
 =1$. If further $\mC_{h,\chi }({\mK}_h^{-1}(S_h (k))) = \mC_{h,\chi }({\mK}_h^{-1}(k))$, then
\begin{equation} \label{g4}
  \mC_{h,\chi }({\mK}_h^{-1}(k))  =
  \frac{1}{\sqrt{\left|det\left(\frac{\partial(\mD_h (S_h(k),q))_\mu}
  {\partial q_\nu}\right)_{k=q}\right|}}.
\end{equation}

\end{document}